\documentclass[twocolumn,preprintnumbers,amsmath,amssymb,superscriptaddress]{revtex4}
\usepackage{graphicx}
\usepackage{dcolumn}
\usepackage{amsmath}

\usepackage{bm}
\def\bea {\begin{eqnarray}}
\def\eea {\end{eqnarray}}

\def\be {\begin{equation}}
\def\ee {\end{equation}}

\begin{document}

\title{Centrality dependent long-range angular correlations of intermediate-$p_{T}$ D-mesons and charged particles in $pPb$ collisions at the LHC energy}

\author{\it Somnath Kar}
\address{Key Laboratory of Quark and Lepton Physics (MOE) and Institute of Particle Physics, Central China Normal University, Wuhan 430079, China}
\author{\it Subikash Choudhury}
\address{Key Laboratory of Nuclear Physics and Ion-beam Application (MOE) and Institute of Modern Physics, Fudan University, Shanghai 200433, China}
\author{\it Samrangy Sadhu}
\author{\it Premomoy Ghosh}
\email{prem@vecc.gov.in}
\address{Variable Energy Cyclotron Centre, HBNI, 1/AF Bidhan Nagar, Kolkata 700 064, India}
        
\begin{abstract}
The high-multiplicity events of $pPb$ collisions at $\sqrt s_{NN}$ = 5.02 TeV at the LHC exhibit unforeseen collective behaviour. One of the possible explanations to 
the collectivity could be the formation of thermalized partonic matter, like the one formed in relativistic nucleus-nucleus collisions and is described by the hydrodynamic 
models. This article presents a study on the centrality dependent long-range two-particle azimuthal correlations of D-mesons and charged particles in $pPb$ collisions 
at $\sqrt s_{NN}$ = 5.02 TeV. The study has been conducted on the events, generated with the EPOS3 hydrodynamic code that reproduces most of the features of the 
$pPb$ data at the LHC energy. There appears a ridge-like structure in the long-range two-particle angular correlations of D-mesons, in the intermediate $p_{T}$-range, 
and charged particles in the simulated high-multiplicity $pPb$ events.\\

\end{abstract}
\maketitle

\section{Introduction}
\label{}
The thermalized partonic matter, the Quark-Gluon Plasma (QGP) \cite {ref01, ref02}, has been observed in experiments \cite {ref03, ref04, ref05, ref06} of ultra-relativistic 
gold-gold ($AuAu$) collisions at the centre-of-mass energy ($\sqrt s_{NN}$) of 130 and 200 GeV at the Relativistic Heavy Ion Collider (RHIC) at the BNL. Prior to the 
discovery at the RHIC, there had been efforts in search of the QGP in relativistic heavy-ion collisions at lower  \cite {ref07, ref08}  $\sqrt s_{NN}$ and even in proton-proton 
($pp$) \cite {ref09, ref10, ref11} collisions. Lack of confirmative signals for the QGP in the lower energy data pushed the requirement of the energy of collisions continually 
upwards. On the other hand, the relativistic heavy-ion collisions, considered to be more conducive to the QGP-thermalization because of the larger volume, longer lifetime 
and involvement of large number of nucleons, became the system of choice in search of the QGP. In extracting the signals of the QGP in the relativistic nucleus-nucleus 
collisions, the proton-proton and proton-nucleus collisions, however, play the role of the base-lines. Of the most significant features observed in the RHIC data, the collective 
flow of the final state particles produced in the collisions indicates thermalization and the suppression of the high-$p_{T}$ particles or the jets points to the formation of dense 
partonic medium. To derive the true medium effect on high-$p_{T}$ suppression, the heavy-ion data is studied in terms of the nuclear modification factor, $R_{AA}$ \cite 
{ref12}, defined as the ratio of the yields in heavy-ion and $pp$ collisions at the same energy in a given $p_{T}$-bin, normalized with the number of binary nucleon-nucleon 
collisions. The effect of the hot nuclear matter or the QGP formed in heavy-ion collisions is finally estimated by disentangling the cold nuclear matter (CNM) effects \cite 
{ref13, ref14, ref15}, experimentally obtained from the proton-nucleus collisions. At the RHIC, however, the CNM effect was studied with deuteron-gold ($dAu$) collisions 
because of technical difficulties for $pAu$ collisions. \\ 

The Large Hadron Collider (LHC) has extended the domain of the QGP study. The heavy-ion program at the LHC experiments with heavier nuclei ($PbPb$) and at 
higher $\sqrt s_{NN}$ (2.76 and 5.02 TeV), create hotter partonic matter with increased energy density, volume and the lifetime \cite {ref16}. It also facilitates the study 
of properties of the medium with copiously produced unique hard probes, the heavy-flavor (HF) particles. The LHC-data also indicate the possibility of formation of the QGP-like 
collective medium in small systems produced in the high multiplicity events of $pp$ \cite {ref17, ref18, ref19, ref20} and $pPb$ \cite {ref21, ref22, ref23, ref24} collisions. 
The recent analysis \cite {ref25} of RHIC data on $dAu$ collisions also corroborate the LHC finding. Though the source of the collectivity in the small systems is not yet 
unambiguously identified, there has been considerable effort in  connecting the novel phenomenon with the QGP-like collectivity. Beside revealing \cite {ref21, ref22, ref23, 
ref24} the long-range two-particle azimuthal angle correlations between charged particles, indicating collectivity in high-multiplicity events of $pPb$ collisions at $\sqrt s_{NN}$ 
= 5.02 TeV, ALICE has studied \cite {ref26} minimum-bias $pPb$ data in terms of two-particle azimuthal correlations between the D-mesons and charged particles, in the 
short-range ($|\Delta\eta| \textless$ 1, where $\eta$ = - $\ln \tan\theta/2$, the pseudorapidity of a particle and $\theta$ is the polar angle of the particle with respect to the 
beam direction). The study reports description of data by the EPOS3 generated events. In this article, we present a multiplicity dependent study of simulated $pPb$ events 
at $\sqrt s_{NN}$ = 5.02 TeV in terms of the two-particle azimuthal correlations between the D-mesons and charged particles, in the long-range (2 $\textless |\Delta\eta| 
\textless$ 4). The analysis has been carried out with the $pPb$ events at $\sqrt s_{NN}$ = 5.02 TeV, generated by the EPOS 3 hydrodynamic model \cite {ref27}, that 
satisfactorily describes the features of collective behaviour of particle production in $pPb$ events at the LHC. \\

\section{Heavy-flavor mesons: the probe}
\label{}
Because of the large masses, the production of HF-quarks (charm and bottom) predominantly takes place in the hard scattering of partons during the primordial stage of
ultra-relativistic heavy-ion collisions. Most of the heavy quarks, produced in the heavy-ion collisions, thus witness the entire evolution of the QGP medium. Also, due to the 
large momentum transfer in the hard partonic interactions, the production cross-sections of heavy-quarks are calculable in the perturbative QCD approach. While diffusing
through the medium, made of the light quarks and gluons, the heavy quarks experience radiative and collisional energy loss that is reflected in the spectra of the final state
HF-mesons. The HF-meson (particularly the D-meson) has already played a significant role in characterising the medium formed in $AuAu$ collisions \cite{ref28, ref29, ref30} 
at RHIC as well as in $PbPb$ \cite {ref31, ref32} and $pPb$ \cite{ref33} collisions at the LHC energies. The high $p_{T}$ charm suppression has been observed in the central 
$AuAu$ \cite {ref28} and $PbPb$ \cite {ref31} collisions. The flow of charm has also been measured \cite {ref29, ref30} in the $AuAu$ collisions at RHIC, in terms of the semi-leptonic 
decayed electrons. The D-mesons have been found to have medium induced collective flow \cite {ref32} in the $PbPb$ collisions at $\sqrt s_{NN}$ = 2.76 TeV. The ALICE has 
measured the nuclear modification factor, $R_{pPb}$, for D-mesons yields \cite {ref33} and the relative yields of D-mesons as a function of relative charged particle multiplicity 
\cite {ref34} in $pPb$ collisions at $\sqrt s_{NN}$ =5.02 TeV. The $R_{pPb}$ measurement \cite {ref33}, revealing very small CNM effects for $p_{T} \geq $ 3 GeV/c, confirmed 
that the suppression of high $p_{T}$ ($\geq $ 2 GeV/c) D-mesons \cite {ref31} in $PbPb$ collisions at $\sqrt s_{NN}$ = 2.76 TeV is predominantly due to final-state effect of the 
charm energy loss in the medium and not due to the initial-state CNM effect. \\

The ``ridge" structure in high-multiplicity events of $pPb$ collisions  \cite {ref21, ref22, ref23, ref24}, as observed in two-particle correlation study with the light-flavored 
particles has been suggested to be due to either collectivity \cite {ref35} or the gluon saturation \cite {ref36}. Though, the collectivity in the high-multiplicity $pPb$ 
events at the LHC energy is largely accepted, this study of long range azimuthal correlations of HF-mesons and charged particles in high-multiplicity $pPb$ events could 
shed further light, in this context. \\

\section {Two-particle angular correlations: the analysis tool}
\label{}

The two-particle angular correlation function is defined by the per-trigger associated yields of charged particles obtained from $\Delta\eta,\Delta\varphi$ distribution 
(where $\Delta\eta$ and $\Delta\varphi$ are the differences in the pseudo-rapidity ($\eta$) and azimuthal angle ($\varphi$) of the two particles) and is given by:\\
\begin{equation}
\frac{1}{N_{trig}}\frac{d^{2} N^{assoc}}{d\Delta\eta d\Delta\varphi} = B(0,0)\times\frac{S(\Delta\eta, \Delta\varphi)}{B(\Delta\eta, \Delta\varphi)}
\label{Eq1}
\end{equation}
where $N_{trig}$ is the number of trigger particles in the specified $p_{T}^{trigger}$ range. \\

The function S($\Delta\eta, \Delta\varphi$) is the differential measure of per-trigger distribution of associated hadrons in the same-event, i.e,\\

\begin{equation}
S(\Delta\eta, \Delta\varphi) = \frac{1}{N_{trig}}\frac{d^{2} N^{assoc}_{same}}{d\Delta\eta d\Delta\varphi}\\
\end{equation}

The same-event distribution functions are corrected for the random combinatorial background and effects due to the limited acceptance by dividing the raw same-event 
distribution function by the mixed-event background distribution, where trigger and associated particles are paired from two different events of similar multiplicity.\\

The background distribution function B($\Delta\eta, \Delta\varphi$) is defined as: 
\begin{equation}
B(\Delta\eta, \Delta\varphi) =\frac{d^{2} N^{mixed}}{d\Delta\eta d\Delta\varphi}
\end{equation}
where $N^{mixed}$ is the number of mixed event pairs.\\

The factor B(0,0) in Eqn.~\ref{Eq1} is used to normalize the mixed-event correlation function such that it is unity at ($\Delta\eta, \Delta\varphi$)=(0,0).
Finally, the acceptance corrected correlation function is determined by scaling the same-event distribution function, $S(\Delta\eta, \Delta\varphi)$ by 
the inverse of the normalized background distribution function, $B(\Delta\eta, \Delta\varphi)/ B(0,0)$.\\

The two-particle azimuthal correlations is a versatile analysis tool that addresses several sources of correlations in multiparticle production, depending on the studied ranges of $|\Delta\eta|$ 
and also the $p_{T}$ for the trigger and the associated particles. The ``short-range" ($|\Delta\eta | \sim 0$) two-particle azimuthal angle correlations are dominated by jets, produced in the hard 
QCD scattering. As the jets are produced back-to-back in azimuth, the jet correlations are reflected in the $|\Delta\varphi |$ - distribution. The jet-induced per trigger hadron-pair yields from the 
same jet populate in the ``near-side" at $|\Delta\varphi | = (|\varphi_{trigger} - \varphi_{assoc.}|) \sim 0$. The pair yields from the ``away-side" jets show up at $|\Delta\varphi | =  (|\varphi_{trigger} 
- \varphi_{assoc.}|) \sim \pi$. On the other hand, a ridge-like structure that appears in the ``long-range" ($|\Delta\eta | \gg 0$) two-particle azimuthal angle correlations in relativistic heavy-ion 
collisions, is attributed to the formation of a collective medium of particle production. The per trigger pair yields with small $|\Delta\varphi |$ over a wide range of $|\Delta\eta |$ (long-range), resulting 
a ``ridge" structure also appears in the high multiplicity $pp$ \cite{ref17, ref18, ref19, ref20} and $pPb$ \cite {ref21, ref22, ref23} events at the LHC. The away-side signal of the long-range 
correlations contains contributions from jet-like correlations also, making it difficult to extract pure signal for long-range correlations. In search of collective medium of particle production, it is the 
near-side ``long-range" or the ``ridge-like" correlations, that becomes important as the near-side structure of the two-particle azimuthal angle correlations in the long-range is considered to be free 
from other effects.  \\

\begin{figure}[htb!]
\centering
\includegraphics[scale=0.45]{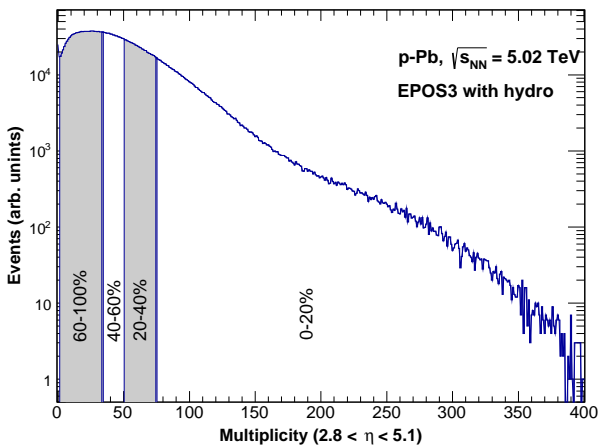}
\caption{Centrality selection for the EPOS3 generated events from the V0A acceptance \cite{ref39} of ALICE set-up.}
\label{fig:V0Amult_EPOS3} 
\end{figure}

\section{Analysis and Results}
\label{}
 
\subsection{Event Generation by EPOS3}
\label{}
The EPOS3 is a hybrid MC event generator having three basic ingredients, a flux-tube initial condition, 3+1D viscous hydrodynamics and a hadronic afterburner modelled via UrQMD. 
In addition, it implements interplay between hard and soft physics processes. The most important aspect of the EPOS3 simulation code for particle production at the LHC energy 
is probably the similar treatment adopted in $proton-proton$, $proton-nucleus$ and $nucleus-nucleus$ collisions, which facilitate understanding the observed feature of collectivity 
in the high-multiplicity $pp$ and $pPb$ events at the LHC vis-a-vis the exhaustively studied collective phenomena in relativistic $nucleus-nucleus$ collisions. \\

In this model, an elementary scattering of partons give rise to parton ladder. Each parton ladder is considered as a longitudinal color field or a flux tubes, carrying transverse 
momentum of the hard scattering. The flux tubes expand and at some stage get fragmented into string segments of quark-antiquark pairs. In case of many elementary parton-parton 
scattering in an event of ultra-relativistic $pp$, $p-nucleus$ or $nucleus-nucleus$ collisions, resulting high-multiplicity, a large number of flux tubes are produced, eventually leading to 
high local string-segment density. The high energy of string segments and / or high local string-segment density (above a critical value) constitute the bulk matter, forming the medium. 
The string segments in the bulk matter, which do not have enough energy to escape, form a ``core" of thermalized ``plasma" that undergoes hydrodynamical expansion following (3 + 1D) 
viscous hydrodynamic evolution followed by by Cooper-Frye mechanisms of particle production. After that, the hadronic evolution takes place till the freeze-out of the ``soft" (low $p_{T}$) 
hadrons. On the other hand, the ``hard" particles or the ``high" $p_{T}$ jet-hadrons originate from hadronization by Schwinger's mechanism of the high-energy string segments from 
the ``corona", the less dense medium in the periphery of the bulk-matter. The ``semi-hard" or the ``intermediate-$p_{T}$-range particles originate from the string-segments with enough 
energy to escape the bulk-matter. These string-segments, while escaping from within the bulk-matter, may pick-up quark-antiquark from the medium. As a result, the intermediate-pt particles 
inherits the properties of the bulk medium. \\

According to the initial conditions of the EPOS3, the heavy quarks may be produced \cite{ref38} in the initial stage, whenever the massive quark - antiquark production is 
possible, through fragmentation of flux tubes or the parton ladders, formed in elementary scattering of partons. In multiple scattering in the EPOS framework, many parton 
ladders are produced, while each of the parton ladders contributes in production of the charm as well as the light quarks leading to the production of D-mesons and light 
hadrons. Although no interaction between the heavy quarks and the bulk thermalized matter is implemented \cite{ref38} in EPOS3, the majority of the particles in the 
``intermediate" $p_{T}$-range in the EPOS framework, which come from the semi-hard string fragmentations, carry the property of the bulk matter and also enough energy 
to escape it. So, the D-mesons originating from the initial semi-hard processes in the intermediate $p_{T}$-range also are likely to reflect the collective property of the bulk 
fluid, like the other light hadrons. With these considerations, to explore the collective nature of the high-multiplicity $pPb$ events at the LHC energy in terms of the long-range 
two particle angular correlations of the D-mesons, in the intermediate $p_{T}$-range, and the charged particles, we have generated 18 million minimum-bias $pPb$ events at 
$\sqrt s_{NN}$ = 5.02 TeV, using the EPOS3.107 code.\\

To make this centrality-dependent study of the simulated events more like data analysis by the experiments, for the centrality estimation, we follow the technique, identical 
to the one followed by the ALICE. Also, we validate the generated events by reproducing the available centrality-dependent ALICE measurements which are relevant to the 
type of analysis we aim to carry out. \\

\subsection{Centrality Estimation}
\label{}

In ALICE, the event classes are obtained either from the signal amplitude in the VZERO detector in the forward / backward rapidity region (2.8 $\textless \eta \textless$ 5.1) 
or from the reconstructed tracklets from the Silicon Pixel Detector (SPD) in the mid-rapidity $|\eta| \textless$ 1.0, in the ALICE experimental set-up. For the centrality 
estimation from the VZERO detector, the minimum-bias events are divided into several event classes, defined as fraction of the analyzed event sample, based on the 
cuts on the total deposited charge in the VZERO detector in the Pb-going direction. The deposited charge on the VZERO detector is proportional to the multiplicity of 
the charged particle in the covered pseudo-rapidity interval. The ALICE measurements show \cite {ref39} that the deposited charge on the VZERO detector or 
equivalently the mean charge particle multiplicity is proportional to the centrality of events. For this analysis with the simulated events, for the centrality selection, we 
consider the charged particle multiplicity in the lead-going direction in the pseudorapidity region of ${2.8 \textless \eta \textless 5.1}$, which is the acceptance of the 
respective VZERO detector in the ALICE set-up. We take into account the asymmetric $pPb$ collisions, where the $nucleon-nucleon$ center-of-mass system moves 
in the direction of the proton beam corresponding to a rapidity of $y_{NN}$ =  - 0.465, resulting the laboratory reference interval $|y_{lab}|  \textless 0.5$ shifting of the 
centre-of-mass rapidity coverage of - 0.96 $\textless y_{cms} \textless$  0.04. In Figure~\ref{fig:V0Amult_EPOS3}, we have shown the fractions of multiplicity distributions 
as the centrality selection for the EPOS3 generated minimum-bias events.\\

\begin{figure}[htb!]
\centering
\includegraphics[scale=0.45]{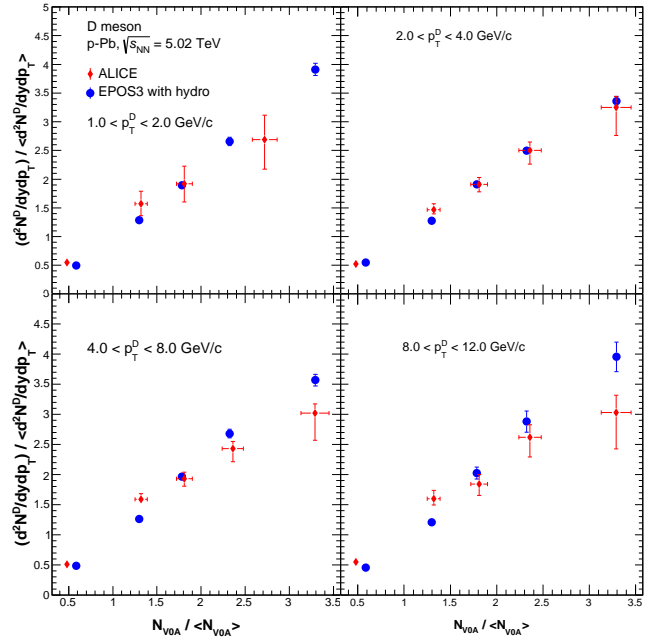}
\caption{Relative D-meson yields as a function of charged particle multiplicity for the $p_{T}$ ranges 1-2, 2 - 4, 4 - 8 and 8 - 12 GeV/c in $pPb$ collisions at $\sqrt s_{NN}$ 
= 5.02 TeV, measured by the ALICE, are compared with the EPOS3 generated events.}
\label{fig:RelativeDmesonYield} 
\end{figure}

\begin{figure}[htb!]
\centering
\includegraphics[scale=0.45]{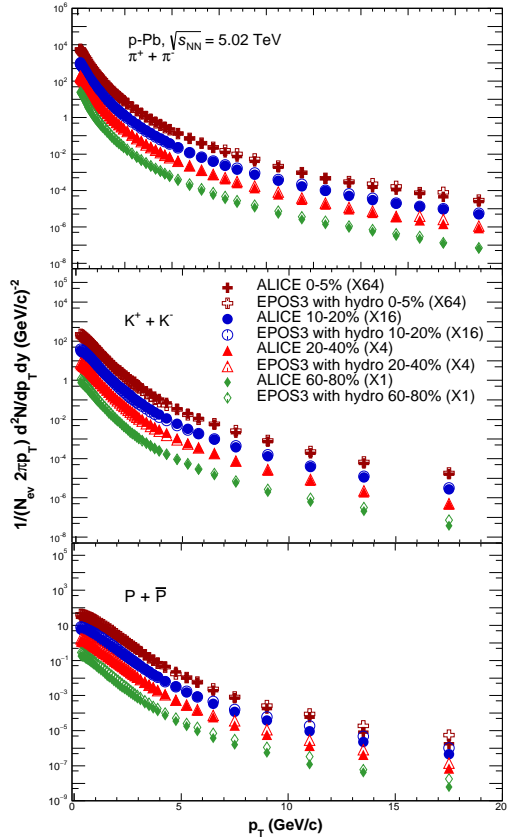}
\caption{Centrality dependent invariant yields of identified charged particles in $pPb$ collisions at $\sqrt s_{NN}$ = 5.02 TeV, measured by ALICE \cite {ref37}, 
are compared with the simulated events from the EPOS3 event generator.}
\label{fig:pTaliceEPOS} 
\end{figure}

\subsection{Validation of generated event-sample}
\subsubsection{Relative yields of D-meson as a function of relative charged particle multiplicity}
\label{}
The ALICE measurement  \cite {ref34} of average relative yields of D-mesons, $(d^2N_{D}/dydp_{T})$ / $\langle d^2N_{D}/dydp_{T} \rangle$ as a function of relative yields of charged particle 
multiplicity $(dN_{ch}/d\eta)$ / $\langle dN_{ch}/d\eta \rangle$ for different $p_{T}$ bins have been reported \cite {ref34} to be well reproduced by EPOS3. To validate our generated events 
to continue with further studies as a function of charged particle multiplicity, we calculate relative yields of D-meson as a function of relative yields of charged particle, $N_{V0A}/<N_{V0A}>$, 
estimated from the pseudorapidity coverage of the acceptance of the VZERO detector of ALICE experiment. In Figure~\ref{fig:RelativeDmesonYield}, we present the centrality or equivalently 
the multiplicity dependence of relative D-mesons yields for four $p_{T}$-bins, 1 to 2. 2 to 4, 4 to 8 and 8 to 12 GeV/c, as obtained from the EPOS3 generated events along with those measured 
by the ALICE. The simulated events reasonably reproduces the measured multiplicity dependence of relative D-meson yields in $pPb$ collisions at $\sqrt s_{NN}$ = 5.02 TeV.\\      

\subsubsection{Centrality dependent invariant yields of identified charged particles}
\label{}
Having the generated events validated by matching the relative D-mesons yields, as measured by ALICE, as a function of relative charged particle multiplicity, it will be relevant to see how the 
generated events describe the measured multiplicity dependent yields of the identified charged particles. The ALICE has measured \cite {ref37, ref40} invariant yields of identified charged particles, 
$\pi^{\pm}$, $K^{\pm}$ and $p$, $\bar p$ for different centrality classes of events. We obtain the invariant yield spectra for the identified charged particles, for  the chosen centrality classes, from 
the EPOS3 generated events and plot the spectra in Figure~\ref{fig:pTaliceEPOS}, along with the respective spectra measured by ALICE. \\

It may be noted that, for comparing with the ALICE data, we have used same scale-factors while plotting the calculated yields in the Figure~\ref{fig:pTaliceEPOS} and because of the chosen scale, 
to accommodate all in one figure, the goodness of the description of the data with the EPOS3 calculations is not clear on visual inspection. Further investigation in terms of ratio, (EPOS3 Calcualtion)/ 
(ALICE Data), reveals that, on an average, the EPOS3 calculated yields of intermediate-$p_{T}$ charged particles lie within about 20 $\%$ of the ALICE data for the top three of the considered centrality
classes. For the event-class of 60-80 $\%$ centrality, the deviation is larger. These calculations are consistent with the previous EPOS3 calculations \cite {ref27} on invariant yields of identified particles 
in $pPb$ collisions.\\  

Our EPOS3-generated event-sample thus reasonably reproduces the multiplicity dependent yields for D-meson and identified charged particles in $pPb$ collisions at $\sqrt s_{NN}$ = 5.02 TeV, 
at least for central events, and the results of calculations are consistent with those of previous EPOS3 calculations \cite {ref27, ref34}. Validating the generated event-sample, we now 
proceed to study the long-range azimuthal correlations between D-mesons and charged particles in high-multiplicity $pPb$ events.\\  

\subsection{Long-range ridge-like correlations}
\label{}
 
In case of formation of collective medium, the long-range two-particle angular correlations of ``soft" particles ideally exist over the entire $|\Delta\eta|$-range. The effect,
however, gets submerged by the dominant jet-like correlation in the short-range ($|\Delta\eta | \sim 0$). On the other hand, the ridge-like, bulk correlations appear prominent 
in the long $|\Delta\eta|$-range, ($|\Delta\eta | \gg 0$) where jet-like short-range correlations are almost absent. At the LHC, ALICE, CMS and ATLAS have studied 
\cite{ref21, ref22, ref23} the centrality dependent long-range two-particle correlations of charged particles in $pPb$ collisions at $\sqrt s_{NN}$ = 5.02 TeV. In this work, for 
the centrality dependent long-range, D-mesons charged particles angular correlations study with the simulated events, we choose the same $|\Delta\eta |$-range and similar 
$p_{T}$-ranges (to start with), as used by the CMS experiment in revealing \cite {ref22} the ridge-like structure in the near-side long-range azimuthal correlations for charged 
particles in the $pPb$ data at $\sqrt s_{NN}$ = 5.02 TeV. This helps us to qualitatively compare our study with existing results from similar analysis, in terms of the $\Delta\varphi$ 
distributions of the per trigger yields. \\

So, for the study of the D-mesons and charged particles angular correlations in the long-range, we consider 2 $\textless |\Delta\eta| \textless$ 4. The CMS experiment 
has studied multiplicity ($N_{track}$) - dependent near-side, long-range angular correlations for charged particles in $pPb$ collisions at $\sqrt s_{NN}$ = 5.02 TeV in 
different $p_{T}$-intervals, 0.1 to 1, 1 to 2, 2 to 3 and 3 to 4 GeV/c, with the same $p_{T}$-ranges for both the triggers and the associated particles. The study revealed 
most prominent ridge-like structure in the high-multiplicity events in the 1 to 2 GeV/c  $p_{T}$-interval. The ridge-like structure diminishes with higher $p_{T}$ and 
nearly disappears in the $p_{T}$-interval 3 to 4 GeV/c.  \\

\begin{figure}[htb!]
\centering
\includegraphics[scale=0.50]{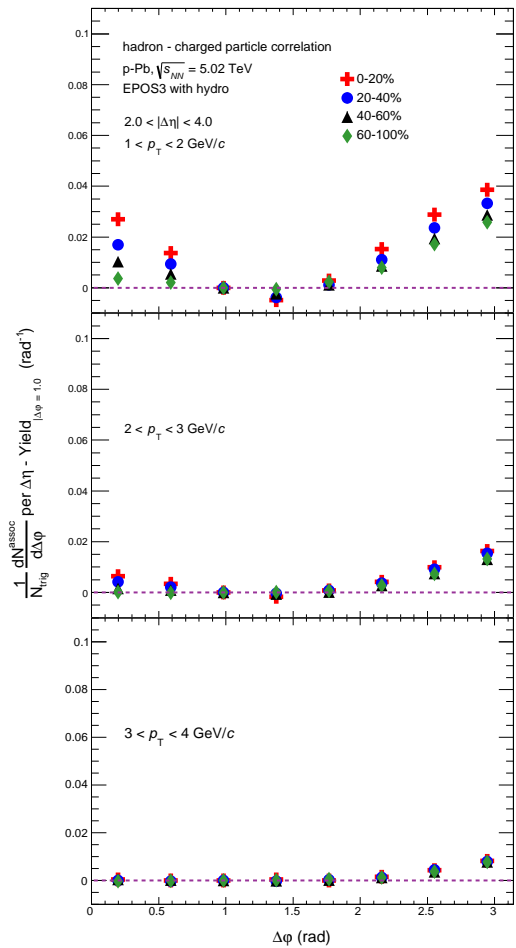}
\caption{Centrality dependent correlated yield as a function $\Delta\varphi$ and subtracted by the $Yield_{|{\Delta\varphi}=1.0}$, as obtained from the long-range two-particle 
azimuthal correlations of hadrons (as triggered particles) and charged particles, averaged over 2 $\textless |\Delta\eta| \textless$ 4, for EPOS3 generated $pPb$ collisions 
at $\sqrt s_{NN}$ = 5.02 TeV in different ranges of the same $p_{T}^{trigger}$ and $p_{T}^{associated}$.}
\label{fig:DelPhiRidgeHCh} 
\end{figure}

We first construct the long-range two-particle azimuthal correlations for the hadrons and the charged particles in the simulated events for the same centrality classes as 
estimated and described in the beginning of this article for the $p_{T}$-intervals 1 to 2, 2 to 3 and 3 to 4 GeV/c. The per trigger correlated yield, projected onto 
$\Delta\varphi$ and subtracted by the $Yield_{|{\Delta\varphi}=1.0}$ (the per trigger correlated yield at ${\Delta\varphi}$ = 1.0) for 2 $\textless |\Delta\eta| \textless$ 4 for 
different centrality bins are obtained and shown in the Figure~\ref{fig:DelPhiRidgeHCh}. The centrality dependence of the correlated yield as a function of $\Delta\varphi$ 
for different $p_{T}$-intervals in the simulated events reveals similar feature as observed in the two-particle azimuthal correlations of the charged particles with the CMS 
data \cite {ref22}: the ridge-like structure is most prominent in the 1 to 2 GeV/c $p_{T}$-range and in the most central events, while it gradually decreases with increasing 
$p_{T}$. \\

\begin{figure}[htb!]
\centering
\includegraphics[scale=0.50]{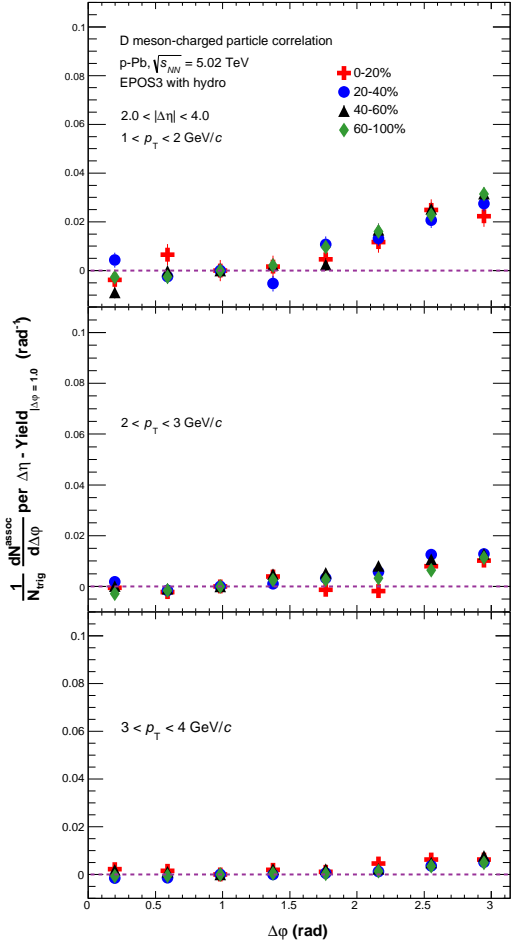}
\caption{Centrality dependent correlated yield as a function $\Delta\varphi$ and subtracted by the $Yield_{|{\Delta\varphi}=1.0}$, as obtained from the long-range two-particle azimuthal 
correlations of D-mesons (as triggered particles) and charged particles, averaged over 2 $\textless |\Delta\eta| \textless$ 4, for EPOS3 generated $pPb$ collisions at $\sqrt s_{NN}$ = 5.02 
TeV in different ranges of the same $p_{T}^{trigger}$ and $p_{T}^{associated}$.}
\label{fig:DelPhiRidgeDCh} 
\end{figure}

Next, the long-range two-particle azimuthal correlations are constructed for D-mesons and charged particles from the simulated events for the same 
centrality classes and in the same $p_{T}$-intervals 1 to 2, 2 to 3 and 3 to 4 GeV/c. The per trigger correlated yields, in the long-range, are projected onto $\Delta\varphi$ 
for different centrality bins. The $\Delta\varphi$ distributions are plotted in the figure~\ref{fig:DelPhiRidgeDCh}. As depicted in the figure~\ref{fig:DelPhiRidgeDCh}, the centrality 
dependence the correlated yield as a function of $\Delta\varphi$ for the different $p_{T}$-intervals in the simulated events in the considered $p_{T}$ intervals do not 
really show the features as observed in case of two-particle correlations of hadrons and charged particles. The non-appearance of the ridge-like structure in the ``low" 
$p_{T}$-range appears consistent in view of the production of the heavy-quarks and their non-interaction with the thermalized bulk-mater in the EPOS3 framework.\\ 

\begin{figure}[htb!]
\centering
\includegraphics[scale=0.45]{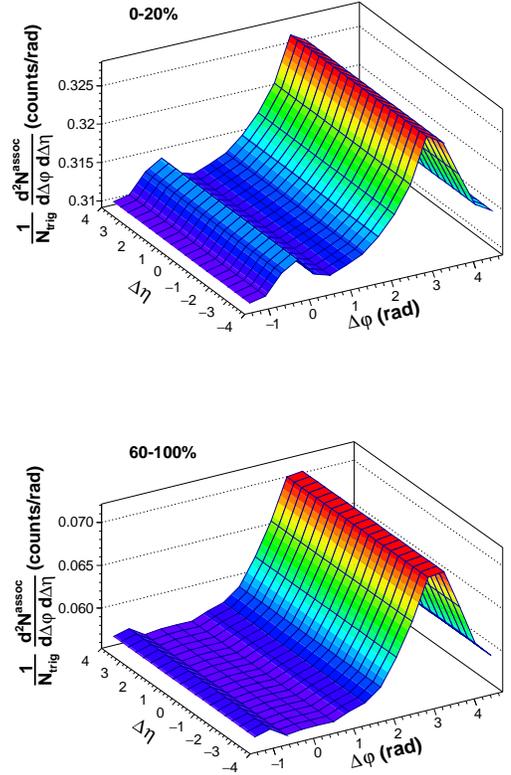}
\caption{Two particle ${\Delta\eta - \Delta\varphi}$ correlation function for 3 $\textless p_{T}^{trigger} \textless$ 5 GeV/c and 1$ \textless  p_{T}^{associated} \textless$ 3 
GeV/c with D-meson as trigger particles for the hydrodynamic-EPOS3 generated $pPb$ collisions at $\sqrt s_{NN}$ = 5.02 TeV in 0 - 20 and 60 - 100 per cent central 
event classes. The short-range correlations has been suppressed for conspicuous presentation of the long-range correlations.} 
\label{fig:DelPhiRidgeDh_3to5_2D} 
\end{figure}

\begin{figure}[htb!]
\centering
\includegraphics[scale=0.45]{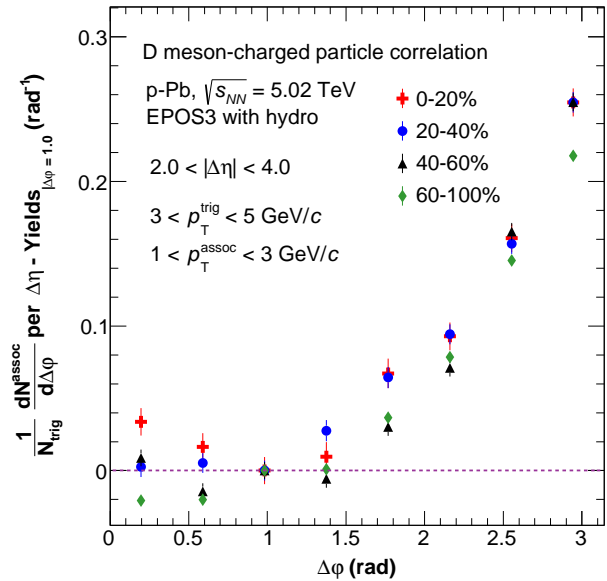}
\caption{Centrality dependent correlated yield as a function $\Delta\varphi$ and subtracted by the $Yield_{|{\Delta\varphi}=1.0}$, as obtained from the long-range 
two-particle azimuthal correlations of D-mesons and charged particles, averaged over 2 $\textless |\Delta\eta| \textless$ 4, for 3 $\textless p_{T}^{trigger} \textless$ 5 
GeV/c and 1$ \textless  p_{T}^{associated} \textless$ 3 GeV/c for the hydrodynamic-EPOS3 generated $pPb$ collisions at $\sqrt s_{NN}$ = 5.02 TeV.}
\label{fig:DelPhiRidgeDh_3to5} 
\end{figure}

At this point, we recollect that ALICE has measured \cite {ref32} significant positive $v_{2}$ (comparable in magnitude to the light-flavored charged hadrons $v_{2}$) of 
the D-mesons in the 2 $\textless p_{T} \textless$ 6 GeV/c range, in 30 - 50 \% centrality class of $PbPb$ collisions at $\sqrt s_{NN}$ = 2.76 TeV. The ALICE result and 
the fact that the measured $v_{2}$ of light charged particles at RHIC and LHC are usually observed to have the positive $v_{2}$ up to the $p_{T}$-range of about 3 GeV/c, 
prompt us to consider respective $p_{T}$-ranges for D-mesons and the charged particles for studying collectivity in terms of the long-range two-particle angular correlations. 
Incidentally and also importantly, as argued in the Section - {\bf IV.A}, the D-mesons in the intermediate $p_{T}$-range, in the hydrodynamic-EPOS3 approach, inherently 
carry the collective property of the bulk fluid. It may also be noted that the modulations in the $\Delta\varphi$ distributions of the two-particle angular correlations actually
represent the cumulative effects due to the $v_{2}$ and it's higher harmonics which, for the long-range correlations, can be factorized as the $v_{n}(p_{T}^{trigger}) v_{n}
(p_{T}^{associated})$. We construct the long-range two-particle azimuthal correlations for the D-mesons and the charged particles for 2 $\textless |\Delta\eta| \textless$ 
4, 3 $\textless p_{T}^{trigger} \textless$ 5 GeV/c and 1$ \textless  p_{T}^{associated} \textless$ 3 GeV/c from 18 million simulated events in the selected centralities. In the 
considered $p_{T}$-ranges, the long-range two-particle azimuthal correlations of D-mesons and charged particles indeed reveal a prominent ridge-like structure in 
the most central event-class, as has been depicted in Figure~\ref{fig:DelPhiRidgeDh_3to5_2D} and Figure~\ref{fig:DelPhiRidgeDh_3to5}, after removing the short-range
jet-like correlations. To compare with the EPOS3 generated events without hydrodynamic evolution we repeat the correlation study on similar statistics of EPOS3 generated 
non-hydro events. The ridge-like structure does not appear for EPOS3 generated high-multiplicity $pPb$ events at $\sqrt s_{NN}$ = 5.02 TeV, without non-hydrodynamic 
evolution, as expected. At this stage, we like to point out that negative values of the correlations which appear in the near-side for the 40 -60 $\%$ and 60 -100 $\%$ centrality 
classes (Figure~\ref{fig:DelPhiRidgeDh_3to5}) have no physics origin and these result as an artefact of baseline subtraction. Since in peripheral collisions, long range region 
may have fluctuations due to few pairs in same event and mixed event which makes determination of baseline uncertain.\\

\begin{figure}[htb!]
\centering
\includegraphics[scale=0.45]{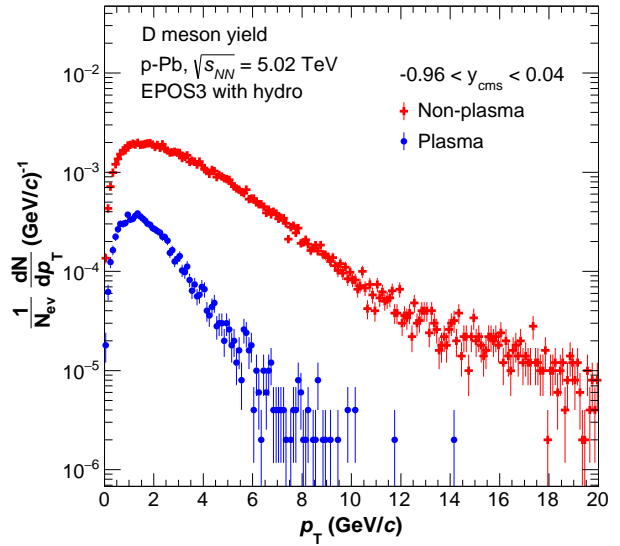}
\caption{Invariant yield of EPOS3 generated D-mesons from two different sources, plasma and non-plasma.} 
\label{fig:yieldSeparateClasses} 
\end{figure}

The appearance of the ridge-like structure in the long-range two-particle angular correlations of the D-mesons, in the intermediate $p_{T}$-range, and  charged particles in the high-multiplicity 
EPOS3 generated $pPb$ events reflects the collective property of the D-mesons. In view of identifying the source of the collective property of the, D-mesons in the EPOS framework, therefore, 
we investigate the generated events further. We select two different classes of D-mesons according to the production mechanisms: 1) "soft" particles from the "core" or the plasma and 2) the 
particles from semi-hard string fragmentation. We calculate the invariant yields separately for the two selected classes of generated particles. The figure~\ref{fig:yieldSeparateClasses} clearly 
shows that the D-meson yield from (semi-)hard string fragmentation (non-plasma source) dominate largely the same from the plasma source in EPOS3 hydrodynamic framework.\\

\begin{figure}[htb!]
\centering
\includegraphics[scale=0.45]{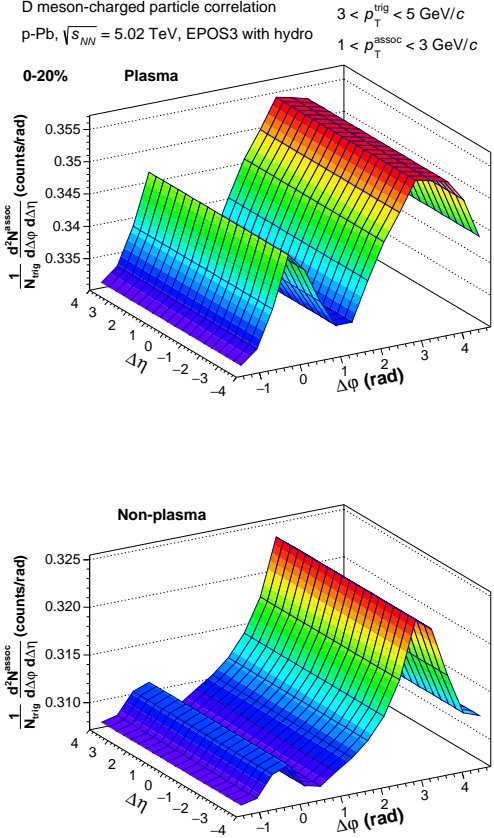}
\caption{Same as Figure~\ref{fig:DelPhiRidgeDh_3to5_2D} for D-mesons generated from the plasma (upper panel) and non-plasma (lower panel) sources for 0 - 20 per cent central 
event class. } 
\label{fig:twoDSeparateClasses1} 
\end{figure}

We calculate the long range two particle angular correlations, between D-mesons from different sources and the charged particles and plot the ${\Delta\eta - \Delta\varphi}$ 
correlation function for 3 $\textless p_{T}^{trigger} \textless$ 5 GeV/c and 1$ \textless  p_{T}^{associated} \textless$ 3 GeV/c with D-meson as trigger particles for the 
hydrodynamic-EPOS3 generated $pPb$ collisions at $\sqrt s_{NN}$ = 5.02 TeV in 0 - 20 and 60 - 100 per cent central event classes, in figure~\ref{fig:twoDSeparateClasses1}
for a) the plasma source and b) the non-plasma source.\\

\begin{figure}[htb!]
\centering
\includegraphics[scale=0.45]{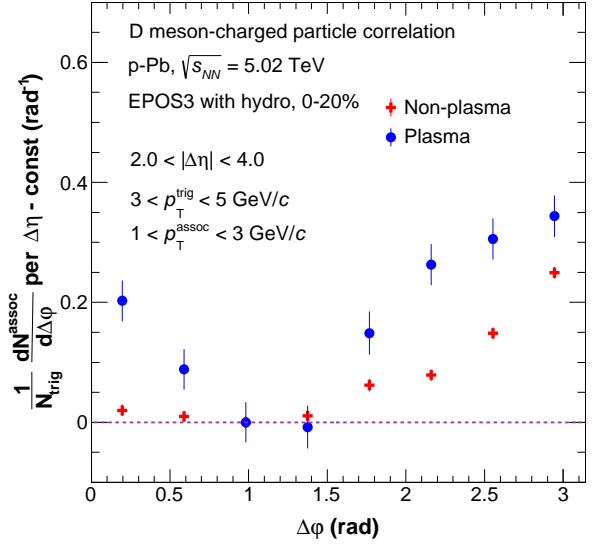}
\caption{Same as Figure~\ref{fig:DelPhiRidgeDh_3to5} for generated particles from two separate sources, plasma and non-plasma.}
\label{fig:oneDSeparateClasses} 
\end{figure}

In the figure~\ref{fig:oneDSeparateClasses}, we compare the centrality dependent per trigger correlated yield from the two sources as a function $\Delta\varphi$ and subtracted by 
the $Yield_{|{\Delta\varphi} =1.0}$, as obtained from the long-range two-particle azimuthal correlations of D-mesons and charged particles, averaged over 2 $\textless |\Delta\eta| 
\textless$ 4, for 3 $\textless p_{T}^{trigger} \textless$ 5 GeV/c and 1$ \textless  p_{T}^{associated} \textless$ 3 GeV/c for the hydrodynamic-EPOS3 generated $pPb$ collisions at 
$\sqrt s_{NN}$ = 5.02 TeV. It becomes clear from the figure~\ref{fig:oneDSeparateClasses}, that the per trigger two-particle correlated yield is more for the D-mesons from the plasma 
source than the one from the non-plasma source. However, as the yield of the D-mesons from the non-plasma source is much more than the yield of that from the plasma source, 
in order of magnitude, the relative contribution of the two sources in the overall per trigger yields become comparable.   \\

\begin{figure}[htb!]
\centering
\includegraphics[scale=0.45]{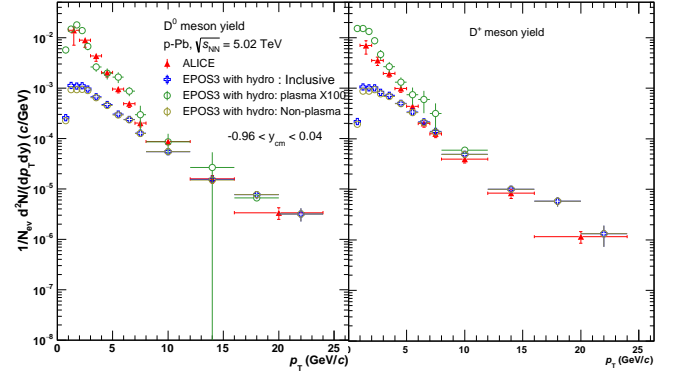}
\caption{$p_{T}$ - differential cross sections of D-mesons in EPOS-hydro generated events, compared with data. The EPOS-yields from plasma is scaled with arbitrary factor to 
compare the the spectral shapes.}
\label{fig:DzeroDplus_Yield_Scaled} 
\end{figure}

At this stage, it will be pertinent to investigate how does the EPOS3 event generator describe the $p_{T}$ - differential cross sections of D-mesons measured at the LHC.
Our analysis reveals (figure~\ref{fig:DzeroDplus_Yield_Scaled}) that the inclusive D-mesons yields match with the ALICE data \cite {ref33} only in the high-$p_{T}$ region, 
$p_{T} >$ 6 GeV/c. On further investigation, we find that D-mesons from hard-scatterings or non-plasma has a reasonable agreement with data in the aforementioned region 
and also it is a dominant source of D-mesons production. We also note that D-mesons from non-plasma, like the inclusive D-mesons, fail to reproduce D-meson yield in 
$p_{T}$ - range, $p_{T} <$ 6 GeV/c. On the other hand, D-mesons production from plasma (green open circle) is highly underestimated, though its spectral shape matches 
with that of the data. It may be noted that the D-meson yields from plasma in EPOS generated events have been scaled with an  arbitrary number and depicted in the 
figure~\ref{fig:DzeroDplus_Yield_Scaled} to compare the spectral shape.\\

\section{Summary and Discussions}
\label{}
We have studied the centrality or the multiplicity dependence of the long-range (2 $\textless |\Delta\eta| \textless$ 4) two-particle angular correlations for the D-mesons and charged particles, 
produced in EPOS3-generated $pPb$ events at $\sqrt s_{NN}$ = 5.02 TeV. This study with EPOS3 generated events is motivated with the observed \cite {ref26} matching of EPOS3 events
and the minimum-bias ALICE data of $pPb$ collisions at $\sqrt s_{NN}$ = 5.02 TeV on the short-range two-particle correlations between D-mesons and charged particles.\\ 

The ridge-like structures as observed \cite {ref22} in the $pPb$ data in the study in two-particle angular correlations between charged particles, in the low $p_{T}$-range (most prominent
in the $p_{T}$ range 1 - 2 GeV/c) is absent in the angular correlations between the D-mesons and the charged particles, in the similar $p_{T}$-range, in the EPOS3 generated events. 
The observation appears to be in accordance with the EPOS3 code, in the present form, where interactions of heavy-quarks with the thermalized bulk matter is not implemented. \\

However, this study on the two-particle angular correlations of D-mesons in the intermediate $p_{T}$-range, (3 $ \textless p_{T}^{trigger} \textless$ 5 GeV/c) and the charged particles  in 
low-$p_{T}$ range, 1 $ \textless p_{T} \textless $ 3 GeV/c, of the EPOS3 generated events clearly shows a prominent ridge-like structure in the long-range in high-multiplicity EPOS3-generated 
$pPb$ events. According to the EPOS3 hydrodynamic model, high-multiplicity events are generated from large number of flux tubes created in many initial parton-parton scatterings in an event. 
A large number of flux-tubes breaks to form a medium of high string-segment density. The low-$p_{T}$ final state particles come from the thermalized bulk-matter created with low energy string-
segments. The semi-hard particles, like the D-mesons in the intermediate $p_{T}$-range, (3 $ \textless p_{T}^{trigger} \textless$ 5 GeV/c), having enough energy to escape the bulk-matter, 
hadronize by picking-up quark or antiquark from the bulk-matter. The D-mesons in this intermediate $p_{T}$-range, thus carry the collective property of the bulk-matter, as reflected in the ridge-like
structure in two-particle angular correlations between the D-mesons in this $p_{T}$ range and charged particles in low-$p_{T}$ range. Further analysis (results depicted in figures 
~\ref{fig:twoDSeparateClasses1} and ~\ref{fig:oneDSeparateClasses}), in terms of correlated pair yields per trigger, suggests that major contribution to the observed ridge-like structure indeed comes 
from the D-mesons produced in the bulk-matter or the ``plasma".  \\

This study addresses the particular issue of formation of collective medium in high-multiplicity $pPb$ collisions in ultra-relativistic collisions and its response to the heavy-flavour particles. The study 
of collectivity and search for its origin in the high-multiplicity $pPb$ events attracted significant attention only after the unexpected experimental observations of collective behaviour in particle production 
in this small system \cite {ref21, ref22, ref23, ref24} at the LHC energies. Moreover, even at the available LHC energies, the statistics of high-multiplicity events $pPb$ collisions is not sufficient yet to 
study the properties of the small collective medium, exhaustively, in terms of all the possible probes, including heavy-flavour particles, the established \cite {ref41, ref42} hard-probe, considered 
to be very effective in characterization of the parton-medium interactions and of the properties of QGP, the strongly interacting matter that is formed \cite {ref03, ref04, ref05, ref06} in ultra-relativistic 
heavy-ion collisions. At this stage, for understanding the collective property of particle production in high-multiplicity $pPb$ collisions, simulation-based studies with well established event generator and 
comparison of $pPb$ data with $PbPb$ data play important roles. In the context of comparing collective property of PbPb and high-multiplicity pPb collisions, a very recent revelation \cite {ref43} 
of an anti-correlation of $v_{2}$ and $v_{3}$ of similar strength for the charged particles in events of same multiplicity of $pPb$ and $PbPb$ data, by strengthening the idea of common origin of the 
collectivity, allows us to discuss existing similar studies, theoretical and experimental, on intermediate-$p_{T}$ heavy flavour particles in $PbPb$ collisions. It is interesting to note that the collective
behaviour of the intermediate $p_{T}$ D-mesons from the EPOS3 generated $pPb$ events at $\sqrt s_{NN}$ = 5.02 TeV is consistent with results from several studies on D-mesons in the similar 
$p_{T}$-range of $PbPb$ collisions data at the LHC. In $PbPb$ collisions at $\sqrt s_{NN}$ = 2.76 TeV the $R_{AA}$ of D-mesons and light-flavour hadrons have been found \cite{ref44} to be 
consistent for $p_{T} >$ 6 GeV/c whereas for $p_{T} <$ 6 GeV/c, the $R_{AA}$ of D-mesons tends to be slightly higher than that of pions. It is worth mentioning that a hybrid model \cite{ref45} of 
fragmentation and coalescence, by incorporating nuclear shadowing effect in the initial state and including both the radiative and collisional energy loss of heavy quarks inside the QGP matter,
can satisfactorily describe the D-meson $R_{AA}$ in central $PbPb$ collisions at $\sqrt s_{NN}$ = 2.76 TeV data \cite{ref44}. The hybrid model calculations show that of the two hadronization processes
of heavy quarks in the QGP medium, the fragmentation and the heavy-light quark coalescence, while the fragmentation dominates for $p_{T} > $ 8 GeV/c, the coalescence becomes crucial
in explaining the data in intermediate and low $p_{T}$-ranges. Further on experimental results, it may be noted that the intermediate $p_{T}$-range of the D-meson, as considered in this study, falls 
within the $p_{T}$-range (2 $ \textless p_{T} \textless$ 6 GeV/c) of the D-mesons in the $PbPb$ collisions at $\sqrt s_{NN}$ = 5.02 TeV \cite {ref32} which exhibit medium induced hydrodynamic 
collectivity in terms of positive $v_{2}$. In another recent study \cite {ref46}, the magnitude of the azimuthal anisotropy coefficients, $v_{2}$ and $v_{3}$ for the prompt $D^0$-mesons for $p_{T} 
\textless$ 6 GeV/c  the $PbPb$ collisions at $\sqrt s_{NN}$ = 5.02 TeV have been reported. In comparison to measurements for the charged particles, these measurements for the prompt $D^0$-
meson have been found \cite {ref46} to be smaller. As the momentum of the D-mesons, with constituent quarks of unequal masses, is mostly contributed by the heavy charm-quark, the constituent 
light-quark has to have low momentum and low $v_{2}$ \cite {ref47, ref48}, resulting slower development of collective features with the $p_{T}$.  \\

Considering observed collective properties of D-mesons, in the intermediate $p_{T}$-range, in ultra-relativistic heavy-ion collisions and several similarities in features of particle production data of 
$pPb$ and $PbPb$ collisions, the prediction of collective behaviour of intermediate-$p_{T}$ D-mesons in the EPOS3 generated high-multiplicity $pPb$ events, which reproduce other observations 
in $pPb$ data, appears reliable.\\

\section {Acknowledgement}
\label{}

The authors are thankful to Klaus Werner for providing them with the EPOS3 code. SC acknowledges financial support from National Science Foundation of China.\\

\end{document}